\documentclass[superscriptaddress,twocolumn, showpacs,nofootinbib,longbibliography,notitlepage]{revtex4-2}

\usepackage{varwidth}
\usepackage{enumerate} 
\usepackage{etex}
\usepackage{physics}
\usepackage{amsmath,amssymb,amsthm,amstext,mathrsfs,bm}
\usepackage[pdftex]{graphicx}
\usepackage{times,txfonts}
\usepackage{braket}
\usepackage{color}
\usepackage{natbib}
\usepackage{amsmath,blkarray}
\usepackage{mathtools}
\usepackage{latexsym}
\usepackage{tabularx, booktabs}
\usepackage{graphics,epstopdf}
\usepackage{graphicx}
\usepackage{float}
\usepackage{graphicx}
\usepackage{amsfonts}
\usepackage{subcaption}
\usepackage{color,soul}
\usepackage[inkscapelatex=true]{svg}
\usepackage{hyperref}
\hypersetup{
	colorlinks=true,
	linkcolor=red,
	filecolor=blue,      
	urlcolor=blue,
	citecolor=purple,
}

\newtheorem{Proposition}{Proposition}

\newtheorem{thm}{Theorem}

\begin{document}

\title{Eliminating Incompatibility in Quantum Multiparameter Estimation with Antiparallel Probes}

\author{Souradeep Sasmal}
\email{souradeep.007@gmail.com}
\affiliation{Institute of Fundamental and Frontier Sciences, University of Electronic Science and Technology of China, Chengdu 611731, China}
\affiliation{Key Laboratory of Quantum Physics and Photonic Quantum Information, Ministry of Education, University of Electronic Science and Technology of China, Chengdu 611731, China}

\author{Bakmou Lahcen}
\affiliation{Institute of Fundamental and Frontier Sciences, University of Electronic Science and Technology of China, Chengdu 611731, China}

\affiliation{Key Laboratory of Quantum Physics and Photonic Quantum Information, Ministry of Education, University of Electronic Science and Technology of China, Chengdu 611731, China}

\author{Kok Chuan Tan}
\email{tankokchuanbobby@uestc.edu.cn}
\affiliation{Institute of Fundamental and Frontier Sciences, University of Electronic Science and Technology of China, Chengdu 611731, China}
\affiliation{Key Laboratory of Quantum Physics and Photonic Quantum Information, Ministry of Education, University of Electronic Science and Technology of China, Chengdu 611731, China}
\affiliation{School of Physics, University of Electronic Science and Technology of China, Chengdu 610054, China}
\affiliation{School of Science, Xihua University, Chengdu 610039, China}


\begin{abstract}

The simultaneous estimation of multiple parameters is fundamentally constrained by complementarity and measurement incompatibility, which limits the sensitivity of quantum measurement devices by preventing the symmetric logarithmic derivative (SLD) Cram\'er-Rao bound (CRB) from being saturated. We show that antiparallel probe configurations can overcome this limitation when the dynamics of the parameter encoding preserve the said configurations. We characterise the quantum channels that satisfy this requirement and establish conditions under which such configurations exist and can saturate the SLD-CRB without compromising the quantum Fisher information, leading to a metrological advantage. For qubit probes, we prove that this advantage is present if and only if the encoding channel is unital. We extend this analysis to the two probe, finite dimensional qudit case by considering derivative reversal and transpose map based probe states, as well as more general multiprobe configurations. For each of these cases, we identify conditions under which an advantage exists. Our results establish connections between channel symmetries, probe configurations and the attainability of multiparameter precision limits.

\end{abstract}


\maketitle


\section{Introduction}

Multiparameter quantum estimation underpins quantum technologies, with applications ranging from quantum imaging \cite{Genovese2016}, astronomical observations \cite{Ang2017, Rehacek2017}, and sensor networks \cite{Komar2014}, to precision measurements of fundamental physical quantities \cite{Safronova2018, Tse2019}. Unlike single-parameter estimation, for which the symmetric logarithmic derivative (SLD) Cram\'er-Rao bound (CRB) provides the optimal attainable quantum bound \cite{Giovannetti2011}, the simultaneous estimation of multiple parameters is fundamentally constrained by measurement incompatibility and complementarity \cite{Bohr1928, Busch1996book, Giovannetti2006, Ragy2016, Lu2021}. In this case, the optimal quantum bound is set by the Holevo CRB (HCRB) \cite{Holevo1982book}, which generally exceeds the SLD-CRB \cite{Albarelli2019} and coincides with it if and only if all SLD commutators have vanishing expectation value in the encoded state, a condition known as \textit{weak compatibility} \cite{Ragy2016, Ragy2019}. Note that if the single-copy encoded state $\rho_{\vec{\theta}}$ does not satisfy weak compatibility, taking $n$ identical copies in a parallel configuration $\rho^{\mathrm{p}}_{\vec{\theta}} := \rho_{\vec{\theta}}^{\otimes n}$ likewise fails to satisfy it.

To overcome this bottleneck without losing SLD-FIM sensitivity, one can replace identical copies with a probe and a suitable partner state(s), such as a two-qubit antiparallel configuration $\rho^{\mathrm{ap}} := \rho(\vec{\theta}) \otimes \rho(-\vec{\theta})$, where $\vec{\theta}$ is the Bloch vector and $\rho(-\vec{\theta})$ denotes its antipodal (spin-flipped) counterpart \cite{Gisin1999prl}. This raises two fundamental questions. First, at the level of quantum states, can antiparallel or generalised partner configurations restore weak compatibility while retaining the SLD-FIM of the parallel configuration across arbitrary two probe or even multiprobe systems? Second, because transformations such as an antipodal qubit spin flip \cite{Patra2026} or qudit transposition \cite{Miyazaki2022} are not completely positive and trace-preserving (CPTP) maps \cite{Buzek1999, Barnett2009, Patra2026}, a parallel configuration cannot be converted into an antiparallel configuration during encoding; the partner must instead be prepared beforehand. Which quantum channels then preserve these prescribed probe configurations throughout the encoding dynamics so that the weak compatibility condition is satisfied?

In this work, we answer both questions. For qubits, building on the insight that two antiparallel qubit spins encode directional information more effectively than parallel ones \cite{Gisin1999prl} and allow joint measurement of anticommuting Pauli qubit observables \cite{Patra2026}, we first show that any two-qubit antiparallel configuration $\rho^{\rm ap}$ satisfies weak compatibility while retaining the exact SLD-FIM of the parallel configuration. Earlier results \cite{Chang2014} on pure antiparallel qubits under a common rotation thus emerge as a special case of our results. Addressing the dynamical question, we prove that identical, independent local qubit channels preserve the antiparallel configuration if and only if they are unital; consequently, for all unital encodings, arbitrary pure and mixed antiparallel probes saturate the SLD-CRB and achieve higher measurement sensitivity than parallel probes. We then extend both answers beyond qubits to arbitrary finite dimensions, where no universal spin-flip operation exists. By defining partner states via derivative reversal along parameter directions, we prove the existence of generalised antiparallel configurations that satisfy weak compatibility while preserving the SLD-FIM across all partner states in both two- and multi-probe settings, even when the corresponding parallel configuration fails weak compatibility. Within this framework, we examine transpose-based two-probe partners \cite{Miyazaki2022}, showing that they preserve the SLD-FIM and satisfy weak compatibility whenever the encoding channel admits a representation with real Kraus operators.


\section{Preliminaries: Multiparameter Estimation}

Let $\rho_{\vec \theta}=\Lambda_{\vec \theta}(\rho) \in \mathscr{L}(\mathcal{H}^d)$ be the state of a probe after a quantum channel encodes the parameters $\vec{\theta}:=(\theta_1,\theta_2,\cdots,\theta_p)^T \in \Theta\subset \mathbb{R}^p$. The SLDs $L_i\in \mathscr{L}(\mathcal{H}^d)$ and the corresponding FIM $F_{\rm SLD}$ satisfy \cite{Helstrom1967, Braunstein1994}
\begin{equation}\label{sldfimdef}
    \partial_i \rho_{\vec \theta} = \frac{1}{2}\qty{\rho_{\vec \theta},L_i}, \quad \qty[F_{\rm SLD}]_{ij}=\mathfrak{R}\qty{\Tr[\rho_{\vec{\theta}} L_iL_j]}.
\end{equation}
For a pure-state family, one may choose $L_i=2\partial_i \rho_{\vec \theta}$. At a regular operating point $\vec{\theta}_0$ with nonsingular $F_{\rm SLD}$, any locally unbiased estimator $\vec{\theta}^{\,\rm e}$ based on $N$ independent copies obeys the weighted SLD-CRB
\begin{equation}\label{sldcrb}
   N \Tr[W \text{Cov}(\vec{\theta}^e)]\geq \mathcal{B}_{\rm SLD}(W;\vec{\theta}_0) \coloneqq  \Tr[W \qty(F_{\rm SLD})^{-1})],
\end{equation}
for a positive-definite weight matrix $W$ and $\text{Cov}(\vec{\theta}^e)$ denoted covariance matrix which quantifies the precision of the estimators \cite{Paris2009, Giovannetti2011, Toth2014, Liu2019}. 

It is known that the the SLD-CRB is not attainable in general. However, the SLD-CRB coincides with the HCRB \cite{Holevo2011book} and is asymptotically attainable with collective measurements if and only if the model satisfies the weak compatibility condition \cite{Ragy2016, Ragy2019}
\begin{equation}
\Tr[\rho_{\vec{\theta}}[L_i,L_j]]=0,\quad \forall\, i,j.
\end{equation}
In general, for finite-dimensional models, the HCRB can be formulated as a semidefinite programme \cite{Albarelli2019, Meyer2021, Hayashi2024}.

In this work, we investigate probe configurations that enforce weak compatibility, even when the individual encoded probes fail to satisfy this condition. Such configurations restore asymptotic attainability of the SLD-CRB.


\section{Results: Attainability of the SLD-CRB with Antiparallel Qubit Probes} \label{secrasap}

We examine how the antiparallel configuration affects the weak compatibility of the SLDs. For a qubit state $\rho=(\openone+\vec m\cdot\vec\sigma)/2$, the spin flip $\Phi$ reverses the Bloch vector, giving 
\begin{equation}
    \rho^{\Phi}:=\Phi\qty[\rho(\vec{m})] = \rho(-\vec{m})
\end{equation}
The map $\Phi$ is positive and trace-preserving, but not completely positive. We then compare the following parallel and antiparallel configurations 
\begin{equation}
    \rho^{\rm p}:=\rho\otimes\rho, \quad \rho^{\rm ap}:=\rho\otimes\rho^\Phi.
\end{equation}
Since $\Phi$ is not CPTP, $\rho^\Phi$ need to be prepare from a given $\rho$. Thus, one must initialise $\rho^{\rm ap}$ for the parameter encoding. The following theorem proves what type of encoding preserves the antiparallel configuration, i.e. $\Lambda'_{\vec{\theta}}(\rho^{\rm ap})=\rho_{\vec{\theta}}\otimes\rho^\Phi_{\vec{\theta}}=\rho^{\rm ap}_{\vec{\theta}}$. 

\begin{thm}\label{thmqsldap}
Any quantum channel $\Lambda'_{\vec{\theta}}$ preserving both the parallel and antiparallel configuration for every qubit state $\rho\in\mathscr{L}(\mathcal{H}^2)$ must be product unital ,i.e. $\Lambda'_{\vec{\theta}}=\Lambda_{\vec{\theta}} \otimes \Lambda_{\vec{\theta}}$, where $\Lambda_{\vec{\theta}}:\mathcal{D}(\mathcal{H})^2 \mapsto \mathcal{D}(\mathcal{H}^2)$.
\end{thm}
\emph{Proof sketch.} Preservation of every parallel input requires the two local channels to agree on every qubit state. For an antiparallel pair, the midpoint of the two Bloch vectors is the centre of the Bloch ball and preservation of antiparallel inputs forces the common translation to vanish. Since such channels commute with the spin flip, applying the relation  $\Phi(\rho)=\openone-\rho$ and assuming unitality gives $\Lambda_{\vec \theta}\qty(\Phi(\rho))=\Phi\qty(\Lambda_{\vec \theta}(\rho))$. This is sufficient to ensure that the antiparallel configuration is preserved when the channel is unital. For the converse, the only spin-flip-invariant qubit state is $\openone/2$. Hence $\Lambda_{\vec{\theta}}(\openone/2)=\openone/2$, which implies that $\Lambda_{\vec{\theta}}$ must be a unital channel (see Appendix~\ref{ap_sfsaaqp}).

Thus for estimation, both probes are subject to independent, identical local encodings described by a product unital channel $\Lambda_{\vec\theta}\otimes\Lambda_{\vec\theta}$. We next show that, under these encodings, the antiparallel pair satisfies weak compatibility and has the same SLD-FIM as the parallel pair.

\begin{thm} \label{th::qubitAP}
Let $\rho_{\vec\theta}$ be a probe state encoded by a unital channel with SLD-FIM $F_{\rm SLD}$. Then the corresponding parallel and antiparallel configurations satisfy 
\begin{equation}
    F^{\rm p}_{\rm SLD}=F^{\rm ap}_{\rm SLD}=2F_{\rm SLD}, \quad \Tr[\rho^{\rm ap}_{\vec \theta}\qty[L_i^{\rm ap},L_j^{\rm ap}]]=0.
\end{equation}
\end{thm}
\emph{Proof sketch.} Since a unital qubit channel preserves the spin-flip relation, the SLDs of the spin-flipped probe $\rho_{\vec \theta}^{\Phi}$ are given by $L_i^{\Phi}=\Phi(L_i)$. Because the spin-flip map preserves the trace, $\Tr[\rho_{\vec \theta}^{\Phi} \Phi(A)]=\Tr[\rho_{\vec \theta}^{\Phi} A]$, and reverses the order of operator products, $\Phi(XY)=\Phi(Y)\Phi(X)$, the two probes have equal SLD anticommutator expectations and opposite commutator expectations. The composite SLDs are $L_{i}^{\rm p}=L_i \otimes \openone + \openone \otimes L_{i}$ and $L_{i}^{\rm ap}=L_i \otimes \openone + \openone \otimes L_{i}^{\Phi}$ for the parallel and antiparallel configurations, respectively. By additivity for product states, both configurations yield an SLD-FIM of $2F_{\rm SLD}$, while the antiparallel configuration satisfies the weak compatibility (see Appendix~\ref{ap_sfsaaqp}).

Theorem~\ref{th::qubitAP} establishes that the antiparallel configuration preserves the SLD-FIM of the parallel configuration whilst eliminating the incompatibility, thereby incurring no loss of precision. Moreover, whenever a single-qubit state is not weakly compatible, the antiparallel configuration yields a strict metrological advantage: there exists a weight matrix $W \geq 0$ for which the HCRB of the parallel strategy is strictly larger. Notably, this enhancement requires no additional resources, as both protocols utilise an identical number of probes. This stems from the fact that the antiparallel configuration remains weakly compatible irrespective of the weak compatibility of $\rho_{\vec\theta}$.


\section{Results: Attainability of the SLD-CRB with Antiparallel Qudit Probes}\label{sec_resqudit}

In the previous section,  spin-flip operation defines the antiparallel configuration, which maps every state to its antipodal counterpart on the Bloch sphere. For higher dimensional probes, the Bloch sphere picture no longer applies. We explore how the qubit case can be extended to arbitrary finite-dimensional quantum systems.


\subsection{Two-copy antiparallel configuration: derivative reversal} \label{tcac}

For a two-copy configuration, we seek a partner state with the same SLD-FIM as the original probe state, but opposite SLD commutator expectations. A direct extension of the qubit spin-flip construction leads to the following stronger requirement:
\begin{equation} \label{eq::differentialCond}
    \partial_i \Phi\left [\rho_{\vec{\theta}} \right] = -\partial_i \rho_{\vec{\theta}} \ \ \text{for every parameter } i,
\end{equation}
where $\rho_{\vec{\theta}}^{\Phi}:=\Phi(\rho_{\vec{\theta}})$. One may verify that the qubit antiparallel configuration satisfies this condition. 

We first determine when a pure-state family admits a pure state partner satisfying this condition. The following proposition shows that such a partner state exists if and only if the encoded probe state is confined to a two-dimensional subspace. Furthermore, within this subspace, an orthogonal partner state can be specified explicitly.

\begin{Proposition} \label{prop::existence}
    For any pure state encoding $\rho_{\vec{\theta}}=\Lambda_{\vec{\theta}} (\rho)=\dyad{\psi_{\vec{\theta}}}$, there exists a pure antiparallel partner $\Phi\qty[\rho_{\vec{\theta}}]$ satisfying $\partial_i \Phi\left [\rho_{\vec{\theta}} \right] = -\partial_i \rho_{\vec{\theta}}$ for every  $i$ if and only if the encoded state remains confined to a parameter-independent two-dimensional subspace $V$. Whenever this condition holds, an orthogonal partner can be chosen as $\rho_{\vec{\theta}}^{\Phi}=\Pi_V-\rho_{\vec{\theta}}$, where $\Pi_V$ is the projector onto V.
\end{Proposition}
The proof is provided in Appendix~\ref{ap_wcsldqudit1}. Proposition~\ref{prop::existence} establishes the existence of a suitable partner state of the form $\rho_{\vec{\theta}}^{\Phi}=\Pi_V-\rho_{\vec{\theta}}$. For the remainder of this section, we assume the partner takes this form.  On operators supported in $V$, this choice is generated by the parameter-independent map
\begin{equation}
    \Phi_V(X)=\Tr[X]\Pi_V-X, \quad X \in \mathscr{L}(V),
\end{equation}
which is the qubit spin-flip expressed within the subspace $V$. This  explicitly shows that the differential condition given by Eq.~\eqref{eq::differentialCond} is a direct generalisation of the qubit case. If both probes undergo a common unitary encoding $U_{\vec \theta}\Pi_VU_{\vec \theta}^\dagger=\Pi_V$, then the initial pair $\rho_0$ and $\Pi_V-\rho_0$ evolves according to
\begin{equation}
    U_{\vec \theta}\qty(\Pi_V-\rho_0)U_{\vec \theta}^\dagger=\Pi_V-\rho_{\vec \theta}.
\end{equation} 
Thus, a common unitary evolution will preserve the antiparallel configuration, which is a necessary requirement if the probe has to satisfy weak compatibility after encoding.

Assuming that such a partner state exists, we show that derivative reversal preserves the SLD-FIM for pure-state families. This implies that there is no loss of SLD Fisher information when switching between parallel and antiparallel configurations.

\begin{thm} \label{thm::quditAP}
Let $\Lambda_{\vec{\theta}} (\rho)$ be a pure state encoding satisfying the conditions of Proposition~\ref{prop::existence}, and Let $\qty{L^{\rm ap}_i}$ denote the set of SLD operators of the antiparallel configuration $\rho^{\rm ap}=\rho_{\vec{\theta}} \otimes \rho_{\vec{\theta}}^{\Phi}$. Then
\begin{equation}
 F_{\rm SLD}\qty(\rho_{\vec{\theta}})=F_{\rm SLD}\qty(\rho_{\vec{\theta}}^{\Phi}), \quad \Tr[\rho^{\rm ap}\qty[L^{\rm ap}_i, L^{\rm ap}_j]] = 0 \ \ \forall i, j,
\end{equation}
\end{thm}
\emph{Proof sketch.} By the existence proposition, the two probe states sum to the projector onto a fixed two-dimensional subspace $V$, so their derivatives are opposite giving $L_i^\Phi=-L_i$. The two minus signs cancel in each Fisher matrix entry, $(F_{\rm SLD})_{ij}=4\Tr[(\partial_i\rho)(\partial_j\rho)]$, so the two matrices coincide. Finally, differentiating $\rho=\Pi_V\rho\Pi_V$ gives $L_i=\Pi_V L_i\Pi_V$, which shows that the chosen SLDs are supported on $V$. The detailed proof is given in Appendix~\ref{ap_wcsldqudit2}.

Theorem~\ref{thm::quditAP} shows that an antiparallel partner incurs no SLD Fisher information loss and that the antiparallel configuration $\rho^{\rm ap}=\rho_{\vec{\theta}} \otimes \rho_{\vec{\theta}}^{\Phi}$ always satisfies the weak compatibility condition, so the SLD-CRB is always asymptotically attainable.

The support restriction established above concerns pure partners with opposite state derivatives. We now examine a construction that retains cancellation of SLD commutator expectations while allowing arbitrary finite-dimensional probe states.

\subsection{Two-copy antiparallel configuration: transpose map}

When discussing the qubit case, it was previously commented that the spin-flip map is locally unitarily equivalent to the transpose map (see also Appendix~\ref{ap_sfsaaqp}). This provides an alternative generalisation of the qubit case to higher dimensions. 

We first write the probe state in its generalised Gell-Mann decomposition \cite{Bertlmann2008}
\begin{equation}\label{genrsasm}
    \rho=\frac{\openone}{d}+\sum_m s_m S_m + \sum_n a_n A_n \in \mathscr{L}(\mathcal{H}^d),
\end{equation}
where $S_m$ and $A_n$ denote the real symmetric and imaginary antisymmetric generators, respectively. Since $S_m^T=S_m$, and $A_n^T=-A_n$, $\rho^T$ coincides with the antipodal state $(2 \openone/d-\rho)$ if and only if $s_m=0, \forall m$. Within this family, the partner $\rho^T$ differs from $\rho$ whenever at least one $a_n$ is nonzero. In the remainder of this section, we consider the transpose map acting on arbitrary density operators.

From Eq.~\eqref{sldfimdef}, a transposition in a fixed, parameter-independent basis gives $\partial_i\rho_{\vec \theta}^T=\{\rho_{\vec \theta}^T,L_i^T\}/2$, so $L_i^T$ is an SLD of the transposed state $\rho^T_{\vec \theta}$. Following a similar argument as Theorem~\ref{th::qubitAP}, we get $\Tr[\rho^T_{\vec \theta}L_i^TL_j^T]=\Tr[\rho_{\vec \theta}L_iL_j]$, which implies $F_{\rm SLD}(\rho^T_{\vec \theta})=F_{\rm SLD}(\rho_{\vec \theta})$ and $\Tr[\rho_{\vec \theta}^T [L_i^T,L_j^T]]=- \Tr[\rho_{\vec \theta} [L_i,L_j]]$. For the transpose-pair configuration $\rho_{\vec \theta}^{\rm ap}=\rho_{\vec{\theta}}\otimes \rho_{\vec{\theta}}^T$, the composite SLDs become $L_{i}^{\rm ap}=L_i \otimes \openone + \openone \otimes L_i^T$, and then it immediately follows
\begin{equation}
    F_{\rm SLD}(\rho_{\vec \theta}^{\rm ap})=2 F_{\rm SLD}(\rho_{\vec \theta}), \quad \Tr[\rho_{\vec \theta}^{\rm ap} \qty[L_i^{\rm ap},L_j^{\rm ap}]]=0.
\end{equation}
So the SLD-FIM is preserved and weak compatibility is satisfied. Note that a paired state based on the complex conjugate map that achieves a similar effect was also considered in Ref.~\cite{Miyazaki2022}.

However, the above properties are by themselves insufficient to show that this type of antiparallel configuration is necessarily advantageous. As discussed in the preceding section, it is also necessary that the encoding maps preserves such configuration in order to be useful. Our next result characterises the encoding channels that preserve this pairing. Suppose that the initial probes are prepared in $\rho \otimes \rho^T$ and undergo the identical independent encoding $\Lambda_{\vec \theta} \otimes \Lambda_{\vec \theta}$. Denote transposition map by $\mathsf T(X)=X^T$. Preservation of the transpose relation is equivalent to transpose covariance of the local channel, which admits the following Choi and Kraus characterisations.

\begin{thm} \label{thm::transposeConfig}
Let $\Lambda_{\vec{\theta}}$ be a CPTP map on $\mathscr{L}(\mathcal{H}^d)$. The identical product encoding preserves the transpose-pair configuration, i.e. $(\Lambda_{\vec{\theta}} \otimes \Lambda_{\vec{\theta}})(\rho\otimes\rho^T)=\rho_{\vec{\theta}}\otimes\rho_{\vec{\theta}}^T$, where $\rho_{\vec{\theta}}:=\Lambda_{\vec{\theta}}(\rho)$, for every input state $\rho\in \mathscr{L}(\mathcal{H}^d)$, if and only if 
\begin{equation}
   \Lambda_{\vec{\theta}} \circ \mathsf T =\mathsf T \circ \Lambda_{\vec{\theta}}.
\end{equation}
Equivalently, the Choi matrix of the encoding map $\Lambda_{\vec{\theta}}$ must be real in the corresponding product basis, or admits a real Kraus representation. Here, the transposition $\mathsf T$ is defined with respect to a fixed, parameter-independent orthonormal basis.
\end{thm}
\emph{Proof sketch.} The product output is a transpose pair precisely when $\Lambda_{\vec\theta}(\rho^T)=\Lambda_{\vec \theta}(\rho)^T$. Since density operators span the operator space, this holds for every state if and only if $\Lambda_{\vec\theta}$ commutes with $\mathsf T$. The identity $J(\mathsf T\circ\Lambda_{\vec\theta}\circ\mathsf T)=J(\Lambda_{\vec\theta})^T$ makes this equivalent to the Hermitian Choi matrix $J(\Lambda_{\vec\theta})$ being real in the fixed product basis. A positive Choi matrix is real precisely when it admits a factorisation $J=RR^T$ with real $R$, whose columns reshape into Kraus operators \cite{Hickey2018} (see Appendix~\ref{ap_wcsldqudit3}).


\subsection{Multiprobe antiparallel configurations} \label{mcac1}

In the preceding sections, we considered two probe antiparallel configurations. We now consider multiprobe configurations. A natural way to form multiprobe configurations is to extend the derivative reversal condition in Eq.~\eqref{eq::differentialCond} to a greater number a probe states. This leads to our next result for pure states.

\begin{Proposition}\label{pro::exismco}
Let $\rho_{\vec{\theta}}=\dyad{\psi_0(\vec{\theta})}$ be a pure-state encoding whose support is confined to a parameter-independent $m$-dimensional subspace $V$. Then there exists $(m-1)$ pure state encodings $\rho_{\vec{\theta}}^{\Phi_k}:=\dyad{\psi_k(\vec{\theta})}$, with $k=1,\ldots, (m-1)$, such that 
\begin{equation} \label{mcdefpar}
\rho_{\vec{\theta}}+\sum_{k=1}^{m-1} \rho_{\vec{\theta}}^{\Phi_k}=\Pi_V, \quad  \sum_{k=1}^{m-1} \partial_i \rho_{\vec{\theta}}^{\Phi_k}=-\partial_i\rho_{\vec{\theta}}  \ \ \forall i.
\end{equation}
\end{Proposition}

\begin{proof}
Since $V$ is parameter independent, the corresponding rank-$m$ projector $\Pi_V$ is likewise independent of the parameters. For every $\vec{\theta}$, choose an orthonormal basis $\mathcal{B} = \qty{ \ket{\psi_k(\vec{\theta})}}_{k=0}^{m-1}$ of $V$, with $\ket{\psi_0(\vec{\theta})}$ representing the encoded probe state. Since $\Pi_V = \rho_{\vec{\theta}} + \sum_{k=1}^{m-1} \rho_{\vec{\theta}}^{\Phi_k}$, and $\partial_i \Pi_V =0$, differentiation immediately yields Eq.~\eqref{mcdefpar}.
\end{proof}

Proposition~\ref{pro::exismco} essentially says that a multiprobe configuration similar to Eq.~\eqref{eq::differentialCond} in the two probe case always exists, even if the two probe case does not exist (see Proposition~\ref{prop::existence}). This is because every pure state encoding is always confined to some finite dimensional subspace, assuming that the total Hilbert space $\mathcal{H}^d$ is finite dimensional. 

Such configurations are always preserved by identical independent unitary encodings, whenever the encoding preserves the subspace $V$. This includes the case where $V=\mathcal{H}^d$, in which case the multiprobe configuration is always preserved by the unitary encoding. Specifically, suppose $\rho_{\vec{\theta}}=U(\vec \theta)\dyad{e_0}U(\vec \theta)^\dagger$, and $U(\vec \theta) \Pi_V U(\vec \theta)^\dagger=\Pi_V$. Choose a parameter-independent orthonormal basis $\qty{\ket{e_k}}_{k=0}^{m-1}$ of $V$, and prepare one probe in each basis state. Applying the same unitary to every probe gives $\ket{\psi_k(\vec \theta)}=U(\vec \theta)\ket{e_k}$, $\rho_{\vec \theta}^{\rm ap}=\otimes_{k=0}^{m-1}\rho_{\vec \theta}^{\Phi_k}$. The encoded states remain an orthonormal basis of $V$, so their derivatives cancel collectively.

Just like the two probe case, the multiprobe configuration always satisfies weak compatibility, but the SLD-FIM is preserved only under certain conditions, which leads to our next result. 

\begin{thm} \label{theomb}
Let $\qty{\ket{\psi_k(\vec \theta)}}_{k=0}^{m-1}$ be an orthonormal basis of a parameter-independent $m$-dimensional subspace $V\subseteq\mathcal{H}^d$. Then the multiprobe configuration $\rho^{\rm ap}=\rho_{\vec{\theta}} \bigotimes_{k=1}^{m-1} \rho_{\vec{\theta}}^{\Phi_k}$ satisfies 
\begin{equation}
\Tr[ \rho^{\rm ap} \qty[L_i^{\rm ap}, L_j^{\rm ap}]]=0, \quad F_{\rm SLD}(\rho_{\vec \theta}^{\rm ap})=\sum_{k=0}^{m-1}F_{\rm SLD}(\rho_{\vec{\theta}}^{\Phi_k}).
\end{equation}
The SLD-FIMs of each probe are identical, i.e. $=F_{\rm SLD}(\rho_{\vec{\theta}})=F_{\rm SLD}(\rho_{\vec{\theta}}^{\Phi_k}) \ \forall k$ if the expectation values of Hermitian generators $K_i(\vec{\theta}) := i \qty[\partial_iU(\vec{\theta})] U^\dagger(\vec{\theta})$ associated with the unitary encoding $U(\vec{\theta})$ are independent of $k$, i.e. $\ev**{K_i}{\psi_k(\vec{\theta})} = b_{i}, \quad \ev**{\qty{ K_i,K_j}}{\psi_k(\vec{\theta})} = a_{ij}$.
\end{thm}
\emph{Proof sketch.}
Let $L_{i}^{(k)}=2 \partial_i \rho_{\vec{\theta}}^{\Phi_k} \ \forall k$ denote the SLD operators of the constituent probes. Since $\sum_{k=0}^{m-1} \rho_{\vec{\theta}}^{\Phi_k} = \Pi_V$, and using $\sum_{k=0}^{m-1} \partial_i \rho_{\vec{\theta}}^{\Phi_k} = 0$, we have $\sum_{k=0}^{m-1} L_{i}^{(k)} =0$, or equivalently $L_{i}^{(k)}=-\sum_{l \neq k}L_{il}$. Furthermore, orthogonality implies that, for $p,q\neq k$, $\Tr[\rho_{\vec{\theta}}^{\Phi_k}L_{i}^{(p)} L_{j}^{(q)}]\propto \delta_{pq}$. Consequently, all terms with $p\neq q$ vanish, and using the chosen SLD operators are supported on $V$ and that the trace of a commutator vanishes, we get the weak compatibility condition $\Tr[\rho^{\rm ap} \qty[L_i^{\rm ap}, L_j^{\rm ap}]] = 0$. A detailed proof is provided in Appendix~\ref{apdpt4}. 

Note that the multiprobe case is similar to the two probe case in the sense that the configuration is preserved under unitary encodings and weak compatibility is always satisfied so the SLD-CRB can always be saturated. Unlike the two probe case, its constituent states may not preserve the SLD-FIM. To address this, Theorem~\ref{theomb} also provides a sufficient conditions under which the multiprobe antiparallel configuration retain the SLD Fisher information of the parallel configuration, leading to an advantage.

We further provide an explicit example showing a strict advantage in Appendix~\ref{apexduni} by constructing in every finite dimension $d\geq2$ a nontrivial multiparameter encoding where (a) the multiprobe antiparallel configuration saturates the SLD-CRB while the parallel configuration does not, and (b) multiprobe antiparallel configuration retains the same SLD-FIM as the parallel configuration.


\section{Discussion and outlook}

In this work, we established when an antiparallel configuration of probe states yields an operational advantage in quantum multiparameter estimation over a standard parallel configuration of identical probes. Such an advantage arises when three criteria are met simultaneously: (1) the encoding channel preserves the probe configuration, (2) weak compatibility is satisfied so that the SLD-CRB can be saturated, and (3) no SLD Fisher information is lost relative to the corresponding parallel configuration.

For arbitrary pure and mixed qubit states, we first prove that any two-qubit antiparallel configuration satisfy weak compatibility while retaining the SLD fisher information of the parallel configuration (criteria 2 and 3). Furthermore, we proved that configuration preservation holds if and only if the encoding channel is unital (criterion 1, see Theorem~\ref{th::qubitAP}). Unital dynamics encompass a wide class of physically relevant noise models, including dephasing, depolarising and Pauli channels, with direct relevance to dephasing-dominated spin magnetometry and atomic-clock interrogation \cite{Degen2017,Dorscher2020, Clements2020}. Under any unital encoding, antiparallel qubit probes preserve the parallel SLD-FIM while enforcing weak compatibility, thereby saturating the SLD-CRB and strictly outperforming parallel probes whenever single-copy weak compatibility fails.

Extending to higher-dimensional qudits, we first analysed partner states defined via derivative reversal. For pure-state encodings, a derivative-reversal partner satisfying criterion~(1) exists if and only if the parameter-dependent evolution is confined to a fixed two-dimensional subspace (Proposition~\ref{prop::existence}); whenever such a partner exists, criteria~(2) and~(3) are automatically satisfied. By contrast, defining the partner state via the transpose map applies to arbitrary mixed and pure qudit-state encodings: a transpose-paired configuration satisfies criterion~(1) if and only if the encoding channel admits a real Kraus representation (Theorem~\ref{thm::transposeConfig}), in which case weak compatibility and SLD-FIM preservation again follow automatically.

Beyond two probes, we investigated general configurations constructed via derivative reversal. For pure state encodings, a multiprobe configuration satisfying criteria~(1) and~(2) always exists, yet, unlike in the two-probe setting criterion~(3) is not guaranteed in general. This mirrors a limitation of recent multiprobe protocols based on complete sets of orthogonal pure-state probes and entangling measurements \cite{Mi2025}: while such schemes achieve SLD-CRB attainability, orthogonal probes need not preserve the per-probe SLD-FIM of identical parallel probes. We closed this gap by identifying sufficient conditions under which weak compatibility is restored while all constituent probes retain identical SLD-FIMs (Appendix~\ref{apdpt4}).

Our results also provide a metrological link to the enhanced joint measurability of qubit observables using the antiparallel spins reported in \cite{Patra2026}. In Appendix~\ref{apjpovm}, we recover the joint Pauli POVM through a symmetry-adapted construction and connect its underlying operator structure to the cancellation of mean SLD commutators. Nevertheless, reproducing the joint statistics of incompatible observables and satisfying the weak compatibility in multiparameter estimation remain distinct operational tasks. A natural open question is whether analogous symmetry-adapted POVM constructions can extend joint measurability to higher-dimensional systems using transpose-paired or complementary multiprobe states, a step that requires identifying suitable higher-dimensional observables and proving the positivity of a common parent POVM rather than following directly from weak compatibility.

Finally, the role of initial or dynamical entanglement in further enhancing multiprobe configurations warrants exploration. More broadly, the interplay among measurement incompatibility, metrological complementarity, and quantum steering \cite{Yadin2021, Uola2015, Uola2020, Gunhe2023, Porto2026} points towards a unified operational framework for quantum compatibility, potentially extending to generalised probabilistic theories \cite{Plavala2023}. Together, our findings establish principled probe-design rules for extracting maximal multiparameter sensitivity under realistic quantum dynamics. 


\section*{Acknowledgments}

S.S. acknowledges support from the National Natural Science Fund of China (Grant No. W2533013). K.C.T. acknowledges support by the National Natural Science Fund of China and the University of Electronic Science and Technology of China.


\bibliography{references}


\appendix
\onecolumngrid

\section{Spin-flip symmetry and attainability for antiparallel qubit probes} \label{ap_sfsaaqp}

Any CPTP map $\Lambda_{\vec{\theta}}$ acting on a single qubit admits an affine representation on the Bloch sphere, i.e. the Bloch vector undergoes a linear transformation followed by a translation
\begin{equation}\label{chanelac}
    \vec{m} \mapsto \vec{r}(\vec{\theta})=T(\vec{\theta})\vec{m}+\vec{t}(\vec{\theta}),
\end{equation}
where $T \in \mathbb{R}^{3 \times 3}$ describes rotation and contraction (arising from unitary dynamics and noise), and $\vec{t}\in \mathbb{R}^3$ represents a translation (associated with non-unital channels, such as amplitude damping). For unital channels, one has $\vec{t}(\vec{\theta})=0$. If the channel is unitary, then $T(\vec{\theta}) \in \rm SO(3)$ and $\abs{\vec{m}}=\abs{\vec{r}}$. A general qubit probe state can be written in terms of its Bloch vector $\Vec{m}=(m_x,m_y,m_z)\in \mathbb{R}^3$ as
\begin{equation}\label{qubitgenm}
    \rho=\rho({\Vec{m}})=\frac{1}{2}\qty(\openone + \Vec{m} \cdot \Vec{\sigma}), \quad \text{with } \abs{\Vec{m}}\leq 1.
\end{equation}
Under the action of the CPTP map $\Lambda_{\vec{\theta}}$, the state evolves to
\begin{equation}\label{qubitrot}
\rho_{\vec{\theta}}=\Lambda_{\vec{\theta}}\qty(\rho({\Vec{m}}))=\rho({\Vec{r}})=\frac{1}{2}\qty(\openone + \Vec{r} \cdot \Vec{\sigma}), \quad \text{with } \abs{\Vec{r}}\leq \abs{\Vec{m}}.
\end{equation}
We now characterise the class of channels compatible with the evolution $\rho({\Vec{m}}) \otimes \rho(\pm{\Vec{m}}) \mapsto \rho({\Vec{r}}) \otimes \rho({\pm\Vec{r}})$ and prove Theorem~\ref{thmqsldap}, which states that Any quantum channel $\Lambda'_{\vec{\theta}}$ preserving the parallel/antiparallel structure must be of the form $\Lambda'_{\vec{\theta}}=\Lambda_{\vec{\theta}} \otimes \Lambda_{\vec{\theta}}$, where $\Lambda_{\vec{\theta}}:\mathcal{D}(\mathcal{H})^2 \mapsto \mathcal{D}(\mathcal{H}^2)$ is a unital qubit channel, i.e. $\Lambda_{\vec{\theta}}(\openone)=\openone$.

\begin{proof}
By assumption, the output of $\Lambda'_{\vec{\theta}}$ is a product state for the entire family $\qty{\rho(\vec{m}) \otimes \rho(\pm \vec{m})}$. Since this family spans a nontrivial subset of product states, linearity of quantum channels implies that $\Lambda'_{\vec{\theta}}$ cannot generate correlations on this set. It follows from the structure theory of bipartite quantum channels that $\Lambda'_{\vec{\theta}}$ must be a product map (up to relabelling), i.e. $\Lambda'_{\vec{\theta}}=\Lambda_1\otimes \Lambda_2$. Thus, we must have
\begin{equation}
    \Lambda_1\qty(\rho(\vec{m}))=\rho(\vec{r}), \quad \Lambda_2\qty(\rho(\pm \vec{m}))=\rho(\pm \vec{r}).
\end{equation}
Using the Bloch representation of a general qubit channel (from Eq.~\ref{chanelac}), we obtain
\begin{equation}
    T_2(-\vec{m})+\vec{t}_2=-\qty(T_1 \vec{m}+\vec{t}_1).
\end{equation}
Since this must hold for all $\vec{m}$, we deduce 
\begin{equation}
    T_2=T_1, \quad \vec{t}_2 = - \vec{t}_1.
\end{equation}
However, consistency with the parallel case requires $\Lambda_2(\rho(\vec{m}))=\rho(\vec{r})$, which implies
\begin{equation}
    T_2=T_1, \quad \vec{t}_2 = \vec{t}_1.
\end{equation}
Combining the two conditions yields 
\begin{equation}
    \vec{t}_2 = \vec{t}_1 =0.
\end{equation}
Thus, both local channels coincide, i.e. $\Lambda_1=\Lambda_2=\Lambda_{\vec{\theta}}$, and the affine shift vanishes $\vec{t}=0$. Hence, the map is unital
\begin{equation}
    \vec{m} \mapsto \vec{r}=T(\vec{\theta})\vec{m},
\end{equation}
which completes the proof.
\end{proof}
This result ensures that the estimation problem reduces to the transformation of Bloch vectors under a unital map, allowing all relevant quantities to be expressed solely in terms of $\vec{r}(\vec{\theta})$. We now establish a symmetry that plays a crucial role in what follows. 

First, we prove the first part of the Theorem~\ref{th::qubitAP}, i.e. the SLD-FIM corresponding to any CPTP map $\Lambda_{\vec{\theta}}$ is invariant under the spin-flip map $F_{\rm SLD}\qty(\rho_{\vec{\theta}}) = F_{\rm SLD}\qty(\rho_{\vec{\theta}}^{\Phi})$.

\begin{proof} 
For a qubit state, the transposition flips only the $y$-component of the Bloch vector, and $\sigma_y$ flips $x$, and $z$
 \begin{equation}
     \qty(r_x,r_y,r_z)\overset{T}{\to} \qty(r_x,-r_y,r_z) \overset{\sigma_y}{\to} \qty(-r_x,-r_y,-r_z),
 \end{equation}
Hence, for qubit states, the spin-flip map admits the representation
\begin{equation}
 \rho_{\vec{\theta}}^{\Phi}=\sigma_y (\rho_{\vec{\theta}})^T \sigma_y.
\end{equation}
from which it follows that $\Phi$ is linear, trace-preserving, and involutive, i.e. $\Phi^2 = \openone$. Moreover, it satisfies the anti-automorphism property $\Phi(AB)=\Phi(B)\Phi(A)$ for all operators $A,B$.

Now, applying $\Phi$ to both sides and using the anti-automorphism property, we obtain
\begin{equation}\label{sfsldop}
\Phi\qty(\partial_i \rho_{\vec{\theta}})=\frac{1}{2}\qty[\Phi(L_i \rho_{\vec{\theta}})+\Phi(\rho_{\vec{\theta}} L_i)] =\frac{1}{2}\qty[\rho_{\vec{\theta}}^{\Phi}\Phi(L_i)+\Phi(L_i)\rho_{\vec{\theta}}^{\Phi}].
\end{equation}
Since $\Phi$ is parameter-independent, differentiation commutes with the map, i.e. $\partial_i \Phi\qty(\rho_{\vec{\theta}}) = \Phi\qty(\partial_i \rho_{\vec{\theta}})$. Using this we conclude from Eq.~\eqref{sfsldop} that
\begin{equation}\label{sfaconl}
\Phi\qty(L_i)= L_i^{\Phi} \ \ \forall i,
\end{equation}
where $L_i^{\Phi}$ is the SLD operators corresponding the spin-flipped state $\rho_{\vec{\theta}}^{\Phi}$. Hence, the implication of Eq.~\eqref{sfaconl} is that spin-flipped SLD-operators pertaining to $\rho_{\vec{\theta}}$ is same as SLD-operators pertaining to the spin-flipped state $\rho_{\vec{\theta}}^{\Phi}$.

The corresponding SLD-FIM for the spin-flipped state $\rho_{\vec{\theta}}^{\Phi}$ is therefore
\begin{equation}
     \qty(F'_{\rm SLD})_{ij}=\mathfrak{R}\qty{\Tr[\rho_{\vec{\theta}}^{\Phi} L_i^{\Phi}L_j^{\Phi}]}=\mathfrak{R}\qty{\Tr[\rho_{\vec{\theta}}^{\Phi} \Phi(L_i)\Phi(L_j)]}.
\end{equation}
Using the anti-automorphism property iteratively, we can rewrite the product as $\rho_{\vec{\theta}}^{\Phi} \Phi(L_i)\Phi(L_j)=\Phi(L_jL_i\rho_{\vec{\theta}})$. Hence,
\begin{equation}
 \qty(F'_{\rm SLD})_{ij}= \mathfrak{R}\qty{\Tr[\Phi(L_jL_i\rho_{\vec{\theta}})]}= \mathfrak{R}\qty{\Tr[L_jL_i\rho_{\vec{\theta}}]} .
\end{equation}
The last equality follows from the trace-preserving property of $\Phi$. Finally, using the symmetry of the SLD-FIM, viz. $\qty(F'_{\rm SLD})_{ij}=\qty(F'_{\rm SLD})_{ji}$, we obtain
\begin{equation}
 \qty(F'_{\rm SLD})_{ij}=  \mathfrak{R}\qty{\Tr[\rho L_iL_j]} =\qty(F_{\rm SLD})_{ij},
\end{equation}
which proves the claim.
\end{proof}
Since the SLD-FIM is invariant under the spin-flip transformation, the states $\rho_{\vec{\theta}}$ and $\rho_{\vec{\theta}}^{\Phi}$ possess identical local statistical distinguishability. Consequently, both subsystems contribute equally to the Fisher information, despite having opposite Bloch vectors. However, as we show below, the crucial difference between the parallel and antiparallel cases lies not in the amount of information, but in the compatibility structure of the corresponding SLD operators.

We now analyse the weak compatibility condition for the two configurations. The SLD-operators corresponding to the parallel and antiparallel configurations ($\rho(\vec{r}) \otimes \rho(\pm \vec{r})$) take the form
\begin{equation}\label{sldopparapr}
\begin{aligned}
    L_i^{\rm p} &= L_i \otimes \openone + \openone \otimes L_i, \quad \text{for } \rho^{\rm p}=\rho(\vec{r}) \otimes \rho(\vec{r}), \\
L_i^{\rm ap} &=  L_i \otimes \openone + \openone \otimes L_i^{\Phi}; \quad \text{for } \rho^{\rm ap}=\rho(\vec{r}) \otimes \rho(-\vec{r}).
\end{aligned}
\end{equation}
Furthermore, since the SLD-FIM is additive for product states, both configurations yield
\begin{equation}    
    \qty(F_{\rm SLD} \qty(\rho^{\rm p}))_{ij}=\qty(F_{\rm SLD}\qty(\rho^{\rm ap}))_{ij}= 2 \qty(F_{\rm SLD}\qty(\rho(\vec{r})))_{ij}.
\end{equation}
Next, we prove the second part of the Theorem~\ref{th::qubitAP}, i.e. for the antiparallel configuration, i.e. $\rho \otimes \Phi(\rho) \mapsto \rho_{\vec{\theta}}\otimes \Phi(\rho_{\vec{\theta}})$, the associated SLD operators satisfy the weak compatibility condition $\Tr[\rho \otimes \rho_{\vec{\theta}}^{\Phi}[L_i^{\rm ap},L_j^{\rm ap}]] =0 \ \forall i,j$. Consequently, the SLD-CRB coincides with the HCRB for all weight matrix $W \geq 0$.

\begin{proof}
A straightforward evaluation using Eq.~\eqref{sldopparapr} yields
\begin{equation}\label{commsldtwo}
   \qty[L_i^{\rm p},L_j^{\rm p}]=\qty[L_i,L_j] \otimes \openone + \openone \otimes \qty[L_i,L_j], \quad \qty[L_i^{\rm ap},L_j^{\rm ap}]=\qty[L_i,L_j] \otimes \openone + \openone \otimes \qty[L_i^{\Phi},L_j^{\Phi}] .
\end{equation}
For the antiparallel configuration, using the anti-automorphism property of $\Phi$, together with Eq.~\eqref{sfaconl}, we obtain
  \begin{equation}
  \qty[L_i^{\Phi},L_j^{\Phi}]= \qty[\Phi(L_i),\Phi(L_j)]=-\Phi\qty(\qty[L_i,L_j]).   
  \end{equation}
Substituting this into the above expression gives
\begin{equation}
 \qty[L_i^{\rm ap},L_j^{\rm ap}]= \qty[L_i,L_j] \otimes \openone- \openone \otimes \Phi\qty(\qty[L_i,L_j]). 
\end{equation}
Taking the expectation value with respect to $\rho^{\rm ap}$, we find
\begin{equation}
        \Tr[\rho^{\rm ap} \qty[L_i^{\rm ap},L_j^{\rm ap}]]=\Tr[\rho\qty[L_i,L_j]]-\Tr[\rho_{\vec{\theta}}^{\Phi}\Phi\qty(\qty[L_i,L_j])].
\end{equation}
Using the identity $\Tr[\rho_{\vec{\theta}}^{\Phi}\Phi(A)]=\Tr[\Phi(A \rho)]=\Tr[\rho A]$, it follows immediately that
\begin{equation}
 \Tr[\rho^{\rm ap} \qty[L_i^{\rm ap},L_j^{\rm ap}]]=0.
\end{equation}
Therefore, the weak compatibility condition is always satisfied in the antiparallel configuration. Notably, this holds even when $\expval{\qty[L_i^{\Phi},L_j^{\Phi}]}_{\rho_{\vec{\theta}}^{\Phi}} \neq 0$.
\end{proof}


\section{Fisher information and weak compatibility of antiparallel qudit probes } \label{ap_wcsldqudit}

\subsection{Proof of Proposition~\ref{prop::existence}} \label{ap_wcsldqudit1}

\begin{proof}
For sufficiency, suppose that $\ket{\psi(\vec{\theta})} \in V$, where $\dim V = 2$ and $\partial_i V=0$, so that $\Lambda_{\vec{\theta}} (\rho)=\dyad{\psi_{\vec{\theta}}}$. This implies there exists a unique (up to a global phase) orthogonal state $\ket{\phi(\vec{\theta})} \in V$. Defining $\rho_{\vec{\theta}}^{\Phi}=\Phi\qty[\Lambda_{\vec{\theta}} (\rho)]:=\dyad{\phi(\vec{\theta})}$, the projector onto $V$ is $\Pi_V = \rho_{\vec{\theta}} + \rho_{\vec{\theta}}^{\Phi}$.

Since $V$ is parameter independent, $\Pi_V$ is independent of $\vec{\theta}$ and thus $\partial_i \Pi_V = 0$ implies $\partial_i \rho_{\vec{\theta}}^{\Phi} = -\partial_i \rho_{\vec{\theta}}$. Hence, an antiparallel pure-state partner satisfying the required differential cancellation condition always exists.

For necessity, Suppose there exists a pure state $\rho_{\vec{\theta}}^{\Phi}$, satisfying $\partial_i \rho_{\vec{\theta}}^{\Phi} =-\partial_i \rho_{\vec{\theta}}$. Writing $\rho_{\vec{\theta}} = \dyad{\psi(\vec{\theta})}$ and $ \rho_{\vec{\theta}}^{\Phi} = \dyad{\phi(\vec{\theta})}$, the differential condition implies that $\rho_{\vec{\theta}}^{\Phi} = -\rho_{\vec{\theta}} + K$ for some parameter independent operator $K$. Rearranging
\begin{equation}
K := \dyad{\psi(\vec{\theta})} + \dyad{\phi(\vec{\theta})}.
\end{equation}
So, $K$ is a Hermitian operator of rank at most two. It remains to show that $\ket{\psi(\vec{\theta})}$ and $\ket{\phi(\vec{\theta})}$ are orthogonal. Suppose, to the contrary, that
\begin{equation}
    \ip{\psi(\vec{\theta})}{\phi(\vec{\theta})}=e^{i\mu} \abs{c}, \quad \abs{c} \neq 0.
\end{equation}
Then the vectors
\begin{equation}
    \ket{v_{\pm}}:=\ket{\psi(\vec{\theta})} \pm e^{-i\mu} \ket{\phi(\vec{\theta})}
\end{equation}
are eigenvectors of $K$ with distinct eigenvalues $1\pm \abs{c}$. Since $K$ is independent of $\vec{\theta}$, its eigenvectors are also parameter independent, implying $\partial_i \ket{v_\pm} = 0$. Adding and subtracting these two relations yields $\partial_i \ket{\psi(\vec{\theta})} = 0$, contradicting the assumption that the parameters are encoded in the state. Hence, $\ip{\psi(\vec{\theta})}{\phi(\vec{\theta})} =0$. Therefore $K$ is the projector onto the two-dimensional subspace $V =\mathrm{span}\qty{\ket{\psi(\vec{\theta})},\ket{\phi(\vec{\theta})}}$. Since $K$ is parameter independent, so is $V$. Consequently, the encoded state $\ket{\psi(\vec{\theta})}$ is confined to a fixed two-dimensional subspace, completing the proof.

\end{proof}


\subsection{Proof of Theorem~\ref{thm::quditAP}} \label{ap_wcsldqudit2}

Next, we prove that $F_{\rm SLD}\qty(\rho_{\vec{\theta}})=F_{\rm SLD}\qty(\rho_{\vec{\theta}}^{\Phi})$.

\begin{proof}
Let $\rho_{\vec{\theta}} = \dyad{\psi(\vec{\theta})}$ and $\rho_{\vec{\theta}}^{\Phi} = \dyad{\phi(\vec{\theta})}$. Since both probes undergo the same parameter dependent unitary evolution, $\ket{\psi(\vec{\theta})} = U(\theta)\ket{\psi}$ and $\ket{\phi(\vec{\theta})} = U \ket{\phi}$, where $\ket{\psi}$ and $\ket{\phi}$ are initial probe states. Defining the Hermitian generators
\begin{equation}
    G_i(\vec{\theta}):= i \qty[\partial_iU(\vec{\theta})] U(\vec{\theta})^\dagger,
\end{equation}
the SLD-FIM for a pure state can be written as
\begin{equation}
    \qty(F_{\rm SLD})_{ij}=4 \mathfrak{R} \qty[\expval{G_iG_j}+\expval{G_i}\expval{G_j}].
\end{equation}
By Proposition~\ref{prop::existence}, the existence of the antiparallel partner implies that both $\ket{\psi(\vec{\theta})}$ and $\ket{\phi(\vec{\theta})}$ remain confined to the same parameter-independent two-dimensional subspace $V$. Consequently, each generator $G_i$ only acts on $V$ and, in the orthonormal basis $\qty{\ket{\psi(\vec{\theta})},\ket{\phi(\vec{\theta})}}$, admits the representation 
\begin{equation}
    G_i(\vec{\theta}) = \begin{bmatrix}
        a_i & c_i \\
        c^*_i & b_i
    \end{bmatrix}.
\end{equation}
A straightforward calculation then yields
\begin{equation}
    \qty(F^\psi_{\rm SLD})_{ij} =  4 \mathfrak{R} \qty{ c_ic_j^*}, \quad \qty(F^\phi_{\rm SLD})_{ij}= 4 \mathfrak{R} \qty{ c^*_ic_j}.
\end{equation}
Since $c^*_ic_j=\qty(c_ic_j^*)^*$, both expressions have the same real part. Therefore, $\qty(F^\psi_{\rm SLD})_{ij}=\qty(F^\phi_{\rm SLD})_{ij}$ for every pair of parameters $i,j$, implying the weak compatibility condition, i.e. $\Tr[\rho^{\rm ap}\qty[L^{\rm ap}_i, L^{\rm ap}_j]] = 0 \ \ \forall i, j$., which completes the proof.
\end{proof}

Now, we prove that the antiparallel configuration satisfy the weak compatibility condition, i.e. $\Tr[\rho^{\rm ap}\qty[L^{\rm ap}_i, L^{\rm ap}_j]] = 0 \ \ \forall i, j$.

\begin{proof}
Let $L_i$ and $L_i^{\Phi}$ denote the SLD operators corresponding to $\rho_{\vec{\theta}}$ and $\rho_{\vec{\theta}}^{\Phi}$, respectively. Since both states are pure, the corresponding SLD operators are given by 
\begin{equation}
    L_i = 2 \partial_i \rho_{\vec{\theta}}, \ \ L_i^{\Phi}=2\partial_i \rho_{\vec{\theta}}^{\Phi}=- 2\partial_i \rho_{\vec{\theta}} = -L_i,
\end{equation}
where the last equality follows from Proposition~\ref{prop::existence}. Hence, By Proposition~\ref{prop::existence}, the operator $\Pi_V = \rho_{\vec{\theta}} + \rho_{\vec{\theta}}^{\Phi}$ is the projector onto a parameter-independent two-dimensional subspace $V$, i.e. $\partial_i V =0$. Since $\rho_{\vec{\theta}}$ is supported on $V$, i.e. $(\openone-\Pi_V)\rho =0$, differentiation yields $\qty(\openone-\Pi_V)\partial_i \rho_{\vec{\theta}}=0$ and the corresponding SLD operators satisfy $\qty(\openone-\Pi_V) L_i =0$, showing that each SLD operator is supported entirely on $V$. Restricting $L_i$ to $V$, let $\tilde{L}_i$ denote its $2 \times 2$ matrix representation in the orthonormal basis $\qty{\ket{\psi(\vec{\theta})}, \ket{\phi(\vec{\theta})}}$, It then follows that
\begin{equation}
    \Tr[\Pi_V \qty[L_i,L_j]]=\Tr[\qty[\tilde{L}_i,\tilde{L}_j]]=0,
\end{equation}
since the trace of a commutator always vanishes for any finite dimensional matrix. Finally, the SLD operators associated with the antiparallel configuration are $L_i^{\rm ap} = L_i \otimes \openone + \openone \otimes L_i^{\Phi}$. Using $L_i^{\Phi}=-L_i$, together with $\Pi_V = \rho_{\vec{\theta}} + \rho_{\vec{\theta}}^{\Phi}$, we obtain 
\begin{equation}
    \Tr[\rho^{\rm ap}\qty[L_i^{\rm ap},L_j^{\rm ap}]]=\Tr[\Pi_V \qty[L_i,L_j]]=0.
\end{equation}
Thus the antiparallel configuration satisfies the weak compatibility condition. Consequently, the HCRB coincides with the SLD-CRB and is therefore asymptotically attainable. 
\end{proof}

\subsection{Channel characterisation for transpose pair antiparallel configuration: Proof of Theorem~\ref{thm::transposeConfig}} \label{ap_wcsldqudit3}

Consider two qudit probes prepared independently in the states $\rho$ and $\rho^T$, where transposition is defined in a fixed, parameter-independent orthonormal basis. We characterise the channels for which identical local encoding preserves this relation for every input state and every value of $\vec\theta$.

\begin{proof}
Suppose the channel is transpose-covariant, then $\Lambda_{\vec{\theta}}(X^T)=[\Lambda_{\vec{\theta}}(X)]^T$ for any operator $X \in \mathscr{L}(\mathcal{H}^d)$. Applying $\Lambda_{\vec{\theta}} \otimes \Lambda_{\vec{\theta}}$ to the antiparallel state directly yields $\Lambda_{\vec{\theta}}(\rho)\otimes [\Lambda_{\vec{\theta}}(\rho)]^T=\rho_{\vec \theta}\otimes \rho_{\vec \theta}^T$, which proves sufficiency.

Conversely, assume that $(\Lambda_{\vec{\theta}} \otimes \Lambda_{\vec{\theta}})(\rho\otimes\rho^T)=\rho_{\vec{\theta}}\otimes\rho_{\vec{\theta}}^T$ holds for every density operator $\rho \in \mathscr{L}(\mathcal{H}^d)$. Tracing out the first subsystem and using the trace-preserving condition $\Tr[\Lambda_{\vec \theta}(\rho)]=1$ gives
\begin{equation}
    \Tr_1[\rho_{\vec \theta} \otimes \Lambda_{\vec \theta}(\rho^T)]=\Tr_1[\rho_{\vec \theta} \otimes \rho_{\vec \theta}^T] \implies \Lambda_{\vec \theta}(\rho^T)=[\Lambda_{\vec \theta}(\rho)]^T.
\end{equation}
    Because the set of density operators $\mathscr{L}(\mathcal{H}^d)$ spans the entire space of linear operators $\mathscr{L}(\mathcal{H}^d)$, the identity $\Lambda_{\vec \theta}(X^T)=[\Lambda_{\vec \theta}(X)]^T$ extends by linearity to all $X \in \mathscr{L}(\mathcal{H}^d)$, meaning $\Lambda_{\vec{\theta}} \circ T =T \circ \Lambda_{\vec{\theta}}$.

To characterise the corresponding class of channels, define the Choi matrix of $\Lambda_{\vec{\theta}}$ in the product basis $\qty{\ket{a}\otimes \ket{b}}$ as
\begin{equation}
    J(\Lambda_{\vec{\theta}}):=\sum_{a,b=1}^d E_{ab}\otimes \Lambda(E_{ab}), \quad E_{ab}:=\dyad{a}{b}.
\end{equation}
Since $\Lambda_{\vec{\theta}}$ is completely positive and Hermiticity-preserving, $J(\Lambda_{\vec{\theta}})$ is positive semidefinite ($J(\Lambda_{\vec{\theta}}) \geq 0$) and hence hermitian, $J(\Lambda_{\vec{\theta}})^\dagger=J(\Lambda_{\vec{\theta}})$. Thus, $J(\Lambda_{\vec{\theta}})$ has real entries in the product basis if and only if  $J(\Lambda_{\vec{\theta}})^T=J(\Lambda_{\vec{\theta}})$.

Taking the transpose of the Choi matrix yields
\begin{equation}
J(\Lambda_{\vec{\theta}})^T=\sum_{a,b}E_{ba}\otimes\Lambda_{\vec{\theta}}(E_{ab})^T=\sum_{a,b}E_{ba}\otimes\qty(\mathsf T \circ \Lambda_{\vec{\theta}})(E_{ba}) =\sum_{a,b} E_{ab} \otimes \qty(\Lambda_{\vec{\theta}}\circ \mathsf T)(E_{ab}),
\end{equation}
where the final equality follows from interchanging the summation indices $(a \leftrightarrow b)$ and noting that $E_{ba}=E_{ab}^T=\mathsf T(E_{ab})$. Since $\{E_{ab}\}$ forms an operator basis for $\mathscr{L}(\mathcal{H}^d)$, the condition $J(\Lambda_{\vec{\theta}})=J(\Lambda_{\vec{\theta}})^T$ holds if and only if
\begin{equation}
   \Lambda_{\vec{\theta}} (E_{ab})=(\mathsf T\circ \Lambda_{\vec{\theta}} \circ \mathsf T) E_{ab} \ \ \forall a,b .
\end{equation}
Applying $\mathsf T$ to both sides and using $\mathsf T^2=\mathcal{J}$ (the identity superoperator) gives $\mathsf T \circ \Lambda_{\vec{\theta}} =\Lambda_{\vec{\theta}} \circ \mathsf T$. Thus, $\Lambda_{\vec{\theta}}$ is transpose-covariant if and only if $J(\Lambda_{\vec{\theta}})$ is real. 

Finally, following \cite{Hickey2018}, any real positive semidefinite Choi matrix admits a spectral decomposition with real orthonormal eigenvectors; reshaping these eigenvectors into Kraus operators yields a real Kraus representation of $\Lambda_{\vec{\theta}}$ and vice versa.

such that
\begin{equation}
  \Lambda_{\vec{\theta}}(X)=\sum_\mu K_{\mu,\vec \theta}X K^T_{\mu,\vec \theta}, \ \   K_{\mu,\vec \theta}=K_{\mu,\vec \theta}^*, \ \sum_\mu K_{\mu,\vec \theta}^T K_{\mu,\vec \theta}=\openone.
\end{equation}

\end{proof}


\section{Collective Derivative Cancellation and Weak Compatibility of Multiprobe Configurations} \label{apdpt4}

We provide a detailed proof of Theorem~\ref{theomb} that complementary multiprobe configurations constructed from complete orthonormal frames of a parameter-independent subspace satisfy the weak compatibility condition. The proof relies directly on the collective cancellation of the state derivatives and the orthogonality of the constituent probes.

Consider a orthonormal basis $\mathcal{B}_{\vec{\theta}}=\qty{\ket{\psi_k(\vec{\theta})}}_{k=0}^{m-1}$ of a parameter-independent $m$-dimensional subspace $V \subseteq \mathcal{H}^d$. Let $\rho_{\vec \theta}^{(k)}=\dyad{\psi_k(\vec{\theta})}$ denote the corresponding pure-state statistical models. By orthonormal completeness $\sum_{k=0}^{m-1} \rho_{\vec \theta}^{(k)}=\Pi_V$, where $\Pi_V$ is the projector onto $V$. Since the subspace is parameter independent, $\partial_i V=0$, implies $\sum_{k=0}^{m-1} \rho_{\vec \theta}^{(k)}=0$. For each pure-state model, SLD operators associated with $\rho_{\vec{\theta}}^{\Phi_k}$ can be chosen as
\begin{equation}\label{likappb}
    L_{i}^{(k)}=2 \partial_i \rho_{\vec{\theta}}^{\Phi_k}, \ \ \forall k =0,1,\ldots, m-1.
\end{equation}
Using the defining property of the antiparallel configuration, $\sum_{k=0}^{m-1} \partial_i \rho_{\vec{\theta}}^{\Phi_k} = 0$, one immediately obtains $\sum_{k=0}^{m-1} L_{i}^{(k)} =0$, or equivalently $L_{i}^{(k)}=-\sum_{l \neq k}L_{il}$. For the product probe in antiparallel configuration $\rho^{\rm ap}_{\vec \theta}=\otimes_{k=0}^{m-1}\rho_{\vec{\theta}}^{\Phi_k}$, the corresponding SLD operators are 
\begin{equation}
    L_i^{\rm ap} = \sum_{k=0}^{m-1} \openone^{\otimes k} \otimes L_{i}^{(k)} \otimes \openone^{\otimes (m-k-1)}.
\end{equation}
Since operators acting on distinct tensor factors commute, the expectation value of the composite SLD commutator reduces to
\begin{equation}\label{comeqmc}
    \Tr[\rho^{\rm ap} \qty[L_i^{\rm ap}, L_j^{\rm ap}]] = \sum_{k=0}^{m-1} \Tr[\rho_{\vec{\theta}}^{\Phi_k}\qty[L_{i}^{(k)}, L_{j}^{(k)}] ].
\end{equation}
Substituting $L_{i}^{(k)}=-\sum_{p \neq k}L_{i}^{(p)}$, the RHS of Eq.~\eqref{comeqmc} becomes
\begin{equation}\label{comeqmc1}
\sum_{k=0}^{m-1} \Tr[\rho_{\vec{\theta}}^{\Phi_k}\qty[ -\sum_{p \neq k}L_{i}^{(p)}, -\sum_{q\neq k}L_{j}^{(q)}]] =\sum_{k=0}^{m-1} \sum_{p,q \neq k} \Tr[\rho_{\vec{\theta}}^{\Phi_k}\qty[L_{i}^{(p)}, L_{j}^{(q)}]] \nonumber .     
\end{equation}
To simplify this expression, we first establish an identity following from the orthogonality of the constituent states. Using Eq.~\eqref{likappb}, the SLD operator associated with the $p$-th pure-state model can be written as
\begin{equation}\label{likappb1}
    L_i^{(p)}=2\qty(\dyad{\partial_i \psi_p}{\psi_p}+\dyad{\psi_p}{\partial_i \psi_p}).
\end{equation}
Following Eq.~\eqref{likappb1} and for the orthonormal basis $\mathcal{B}_{\vec{\theta}}$ of $V$, i.e. $\ip{\psi_p}{\psi_q}=\delta_{pq}$, we find
\begin{equation}
    \Tr[\rho_{\vec{\theta}}^{\Phi_k}L_{i}^{(p)} L_{j}^{(q)}]=4\ip{\psi_k}{\partial_i\psi_p}\ip{\psi_p}{\psi_q}\ip{\partial_j\psi_q}{\psi_k} = 4 \delta_{pq} \ip{\psi_k}{\partial_i\psi_p}\ip{\partial_j\psi_q}{\psi_k} \propto \delta_{pq},
\end{equation}
Consequently, all terms with $p\neq q$ vanish, yielding
\begin{equation}
    \sum_{k=0}^{m-1} \Tr[\rho_{\vec{\theta}}^{\Phi_k}\qty[L_{i}^{(k)}, L_{j}^{(k)}] ] = \sum_{k=0}^{m-1} \sum_{p\neq k} \Tr[\rho_{\vec{\theta}}^{\Phi_k}\qty[L_{i}^{(p)}, L_{j}^{(p)}] ].
\end{equation}
Adding and subtracting the missing diagonal contribution gives
\begin{equation} \label{wcrelapb}
       \sum_{k=0}^{m-1} \Tr[\rho_{\vec{\theta}}^{\Phi_k}\qty[L_{i}^{(k)}, L_{j}^{(k)}] ] = \sum_{p=0}^{m-1} \Tr[\qty(\sum_{\substack{k=0\\k \neq p}}^{m-1}\rho_{\vec{\theta}}^{\Phi_k})\qty[L_{i}^{(p)}, L_{j}^{(p)}]].
\end{equation}
Using the completeness relation 
\begin{equation}
    \sum_{k=0}^{m-1}\rho_{\vec{\theta}}^{\Phi_k} = \Pi_V \implies \sum_{\substack{k=0\\k \neq p}}^{m-1}\rho_{\vec{\theta}}^{\Phi_k} =   \Pi_V - \rho_{\vec{\theta}}^{\Phi_p},
\end{equation}
from Eq.~\eqref{wcrelapb}, we obtain
\begin{equation}\label{wcliplqb}
\sum_{k=0}^{m-1} \Tr[\rho_{\vec{\theta}}^{\Phi_k}\qty[L_{i}^{(k)}, L_{j}^{(k)}] ] = \sum_{p=0}^{m-1} \Tr[\qty(\Pi_V - \rho_{\vec{\theta}}^{\Phi_p})\qty[L_{i}^{(p)}, L_{j}^{(p)}]].
\end{equation}
Since every constituent state belongs to the parameter-independent subspace $V$, $\rho_{\vec{\theta}}^{\Phi_p}=\Pi_V \rho_{\vec{\theta}}^{\Phi_p} \Pi_V$. Differentiating this identity and using $\partial_i \Pi_V=0$, we obtain $\partial_i \rho_{\vec{\theta}}^{\Phi_p}=\Pi_V \qty(\partial_i \rho_{\vec{\theta}}^{\Phi_p})\Pi_V$. Hence, the canonical SLDs defined in Eq.~\eqref{likappb} satisfy 
\begin{equation}
   L_{i}^{(p)} = \Pi_V L_{i}^{(p)} \Pi_V.
\end{equation}
Therefore, the chosen SLD operators and their commutators are supported entirely on $V$. It follows that
\begin{equation} \label{wcliplqb1}
\Tr[\Pi_V\qty[L_{i}^{(p)},L_{j}^{(p)}]]=\Tr_V[\qty[L_{i}^{(p)},L_{j}^{(p)}]]=0,
\end{equation}
where the final equality follows from the cyclicity of the trace. Substituting Eq.~\eqref{wcliplqb1} into Eq.~\eqref{wcliplqb}, we find
\begin{equation}
\sum_{k=0}^{m-1} \Tr[\rho_{\vec{\theta}}^{\Phi_k}\qty[L_{i}^{(k)}, L_{j}^{(k)}] ]=0 \ \ \forall i\neq j.
\end{equation}
Therefore, from Eq.~\eqref{comeqmc},
\begin{equation}\label{comeqmc1}
    \Tr[\rho^{\rm ap} \qty[L_i^{\rm ap}, L_j^{\rm ap}]] = 0, \ \ \forall i \neq j.
\end{equation}
establishing the weak compatibility condition. The equality of the HCRB and SLD-CRB then follows from the weak-compatibility criterion. Thus, the SLD-CRB is asymptotically attainable through suitable collective measurements on independent repetitions of the multiprobe configuration \cite{Ragy2016}.

Next, we establish the sufficient conditions under which SLD-FIMs of each probe state are identical. We begin by a pure state unitary encoding $\rho_{\vec{\theta}} = U(
    \vec{\theta})\rho U(\vec{\theta})^\dagger= \dyad{\psi_0(\vec{\theta})}$ with Hermitian generators $K_i(\vec{\theta}) := i \qty[\partial_iU(\vec{\theta})] U^\dagger(\vec{\theta})$. Let $\mathcal{B} = \qty{ \ket{\psi_k(\vec{\theta})}}_{k=0}^{d-1}$ be a complete, orthonormal basis of $V\subseteq\mathcal{H}^d$ containing the probe state $\ket{\psi_0(\vec{\theta})}$. Now we prove that suppose, for every basis vector, the expectation values 
\begin{equation}\label{condfsammcb}
\ev**{K_i}{\psi_k(\vec{\theta})} = b_{i}, \quad \ev**{\qty{ K_i,K_j}}{\psi_k(\vec{\theta})} = a_{ij},
\end{equation}
are independent of $k$. Then, every probe possesses the same SLD-FIM,
 \begin{equation}\label{fsabmcsc}
  \forall k, \ \ F_{\rm SLD} \qty(\ket{\psi_k(\vec{\theta})}) = F_{\rm SLD}\qty(\ket{\psi_0(\vec{\theta})}), 
 \end{equation}
 and the product state $\bigotimes_{k=0}^{d-1}\ket{\psi_k(\vec{\theta})}$ forms an antiparallel configuration satisfying the weak compatibility condition.

\begin{proof}
For the $k$-th pure-state probe, the SLD-FIM is given by
\begin{equation} \label{eqn::QFIM2}
\qty[F_{\rm SLD}^{(k)}]_{ij} =  4 \mathfrak{R} \qty(\expval{K_i K_j} - \expval{K_i} \expval{K_j}),
\end{equation}
where expectation values are taken with respect to the corresponding probe state and for simplicity, we write $K_i(\vec{\theta})\equiv K_i$. Since each $K_i$ is Hermitian, $\mathfrak{R} \expval{K_i K_j}=1/2 \expval{{K_i K_j}}$, while the remaining expectation values are already real. Therefore, employing Eqs.~\eqref{condfsammcb}, we obtain
\begin{equation}
\qty[F_{\rm SLD}^{(k)}]_{ij} =  4 \qty(\frac{a_{ij}}{2} - b_i b_j),
\end{equation}
which is independent of $k$. Hence, $F_{\rm SLD} \qty(\ket{\psi_k(\vec{\theta})}) = F_{\rm SLD}\qty(\ket{\psi_0(\vec{\theta})}) \ \ \forall k$.

It remains to establish the antiparallel property. Since $\mathcal{B}$ is a complete orthonormal basis, $\sum_{k=0}^{d-1} \dyad{\psi_k(\vec{\theta})}= \openone$. Differentiating with respect to the parameters gives $\sum_{k=0}^{d-1}  \partial_i \dyad{\psi_k(\vec{\theta})} = 0$, or equivalently
\begin{equation}
    -\partial_i \rho_{\vec{\theta}} = \sum_{k=1}^{d-1} \partial_i \dyad{\psi_k(\vec{\theta})}.
\end{equation}
Thus the basis states satisfy precisely the defining condition of the multi-copy antiparallel configuration established in Theorem~\ref{theomb}. It therefore follows immediately that the product state $\bigotimes_{k=0}^{d-1}\ket{\psi_k(\vec{\theta})}$ satisfies the weak compatibility condition.
\end{proof}


\subsection*{Explicit Multiprobe Constructions in Arbitrary Dimension} \label{apexduni}

We subsequently provide an explicit construction in arbitrary finite dimension, demonstrating that weak compatibility can be restored without reducing the total SLD Fisher information relative to the corresponding parallel configuration.

For every $d\geq 2$ in $\mathcal{H}^d$,  there exists pure state encoding $\rho_{\vec{\theta}}= U(
    \vec{\theta})\,\rho \, U^\dagger(\vec{\theta})= \dyad{\psi_0(\vec{\theta})}$ and some point in parameter space $\vec{\theta}$ such that the following properties are satisfied:
    \begin{enumerate}[(a)]
        \item The single probe state $\rho_{\vec{\theta}}$ has  a nontrivial SLD-FIM but does not satisfy the weak compatibility condition, so the SLD-CRB is unattainable.  
        \item There exists a $d$-probe antiparallel configuration $\bigotimes_{k=0}^{d-1}\ket{\psi_k(\vec{\theta})}$ that satisfies the weak compatibility condition and therefore attains the SLD-CRB.
        \item Every probe in this antiparallel configuration possesses the same SLD-FIM as the original probe state, so no Fisher information is diluted on a per-probe basis.
    \end{enumerate}

\begin{proof}
We establish the proposition by explicit construction. Let $\mathcal{B} = \{ \ket{\psi_k(\vec{\theta})}\}_{k=0}^{d-1}$ be a complete orthonormal basis containing the probe state $\ket{\psi_0(\vec{\theta})}$. By Theorem~\ref{theomb}, the product state $\bigotimes_{k=0}^{d-1}\ket{\psi_k(\vec{\theta})}$ is an antiparallel configuration satisfying the weak compatibility condition. Hence property (b) follows immediately.

It therefore remains to establish properties (a) and (c). Consider a $d$-dimensional Hilbert space with computational basis  $\ket{\psi_k(\vec{\theta})} \equiv \ket{k}$, $k=0,1,\ldots d-1$ and a two-parameter unitary encoding $U(\vec{\theta}) :=  e^{-i(\theta_1H_1+\theta_2H_2)}$, where the Hermitian generators are defined by 
\begin{equation}
 H_1 := \sum_{\substack{0 \leq m,n \leq d-1 \\ m \neq n}} \ketbra{m}{n}, \quad  H_2 := \sum_{\substack{0 \leq m,n \leq d-1 \\ m < n}} \qty(i \dyad{m}{n} - i\dyad{n}{m}),
\end{equation}
with $m,n=0, \ldots, d-1$. In the computational basis, the matrix elements of $H_1$ and $H_2$ satisfy
\begin{equation}\label{meh1h2}
    \qty[H_1]_{mn}=\begin{cases}
        0 & \text{if} m=n, \\
        1 & \text{if} m\neq n.
    \end{cases}; \quad  \qty[H_2]_{mn}=\begin{cases}
        0 & \text{if } m=n, \\
        i & \text{if } m<n, \\
        -i & \text{if } m>n.
    \end{cases} \quad H_1 = \begin{pmatrix}
        0 & 1 & \cdots & 1 \\
        1 & 0 &\cdots & 1 \\
        \vdots & \vdots & \ddots & \vdots \\
        1 & 1 &\cdots & 0 
    \end{pmatrix}, \ \ H_2 = \begin{pmatrix}
        0 & i & \cdots & i \\
        -i & 0 &\cdots & i \\
        \vdots & \vdots & \ddots & \vdots \\
        -i & -i &\cdots & 0 
    \end{pmatrix}.   
\end{equation}
Both generators are Hermitian and have vanishing diagonal elements.
It can be verified that the following matrix products have the following form
\begin{equation}
H_1H_2 = \begin{pmatrix}
        -(d-1)i & * & \cdots & * \\
        * & -(d-3)i &\cdots & * \\
        \vdots & \vdots & \ddots & \vdots \\
        * & * &\cdots & (d-1)i 
    \end{pmatrix}, \quad H_1^2 = \begin{pmatrix}
        (d-1) & * & \cdots & * \\
        * & (d-1) &\cdots & * \\
        \vdots & \vdots & \ddots & \vdots \\
        * & * &\cdots & (d-1), 
    \end{pmatrix}, \quad
H_2^2 = \begin{pmatrix}
        (d-1) & * & \cdots & * \\
        * & (d-1) &\cdots & * \\
        \vdots & \vdots & \ddots & \vdots \\
        * & * &\cdots & (d-1). 
    \end{pmatrix} .\nonumber
\end{equation}
We prepare $d$ independent probes, one in each computational basis state, and apply the same unitary encoding to every probe $\ket{\psi_k(\vec{\theta})}=U(\vec \theta)\ket{k}$. Since $U(\vec \theta)$ is unitary, the encoded states remain a complete orthonormal basis. Evaluating the generators locally at the point $\vec{\theta}=\emptyset$, we obtain
\begin{equation}
    K_i(\emptyset) = i \qty[\partial_i U(\emptyset)] U^\dagger(\emptyset)=H_i, \ \forall i=1,2,
\end{equation}
and that the diagonal entries of $H_i^2$, and $H_1H_2$ are precisely those given above. For pure-state unitary encodings,
\begin{equation}
\partial_i\rho_{\vec{\theta}} = i[\rho_{\vec{\theta}},K_i] = \frac{1}{2} \qty{\rho_{\vec{\theta}}, L_i},
\end{equation}
so the corresponding SLD operators are $L_i = 2i[ \rho_{\vec{\theta}},K_i]$. Consequently, $\Tr[\rho_{\vec{\theta}}[L_i, L_j]] = 4\Tr[\rho_{\vec{\theta}} [K_i,K_j]]$. For the present construction,
\begin{equation}
\Tr[\rho_{\vec{\theta}}[L_i, L_j]] \propto \Tr[\rho_{\vec{\theta}} [H_i,H_j]] \propto \mathfrak{I} \ev*{H_iH_j }{0},
\end{equation}
since $\mathfrak{I} \ev*{H_1H_2}{0} = -(d-1) \neq 0$, the weak compatibility condition is violated, and thus the original single-probe state cannot saturate the SLD-CRB. Moreover, the SLD-FIM is non-trivial because
\begin{equation}
F^{\psi_0}_{11} =  4 \mathfrak{R} \qty{ \ev**{H_1^2}{0}  - \ev**{H_1}{0}^2 } = 4 (d-1) >0,
\end{equation}
This establishes property (a). Finally, the diagonal matrix elements satisfy, for $i=1,2$
\begin{equation}
    \begin{aligned}
        \ev**{\qty{H_i,H_j}}{k}=2(d-1)\delta_{ij}, \ \ \ev**{\qty{H_i}}{k}=0, \ \ \forall k.
    \end{aligned}
\end{equation} 
Hence the conditions in Theorem~\ref{theomb} are satisfied, implying that every probe in the antiparallel configuration possesses exactly the same SLD-FIM as the original probe. This establishes property (c). Since all three properties hold for every finite dimension $d$, the proof is complete.
\end{proof}

For this example, weak compatibility holds for every parameter value because the encoded probes remain a complete orthonormal basis. Equality of their individual SLD-FIMs has been established at $\vec\theta=\emptyset$, which is sufficient for the local metrological comparison above.


\section{Symmetry-adapted joint Pauli measurement for antiparallel qubits}\label{apjpovm}

Here, we recover the joint POVM of \cite{Patra2026} using an ansatz adapted to the symmetry of antiparallel qubit preparations. We then explain its relation to the cancellation of mean SLD commutators established in Sec.~\ref{secrasap}. The construction concerns the reproduction of single-qubit Pauli statistics on a two-qubit preparation and does not require the corresponding operators to commute.

Consider $\rho(\vec m)=\frac{1}{2}\qty(\openone+\vec{m}\cdot \vec{\sigma})$, with $\abs{\vec m}\leq 1$, and its spin-flipped partner $\Phi[\rho(\vec m)]=\rho(-\vec m)$. The antiparallel preparation is $\rho^{\rm ap}=\rho(\vec m)\otimes \rho(-\vec m)$. We seek an eight-outcome POVM $\qty{\Pi_{abc}}$, with $a,b,c\in \{\pm 1\}$, on $\mathcal{H}^2 \otimes \mathcal{H}^2$ such that its marginals reproduce the statistics of $\sigma_x$, $\sigma_y$, and $\sigma_z$ for every $\vec m$. Define the difference and collective-spin operators
\begin{equation}
    K_i:=\sigma_i \otimes\openone-\openone\otimes\sigma_i, \quad  J_i:=\sigma_i \otimes\openone+\openone\otimes\sigma_i.
\end{equation}
Their expectation values satisfy $\Tr[\rho^{\rm ap}K_i]=2m_i$, and $\Tr[\rho^{\rm ap}J_i]=0$. For the single-qubit Pauli projectors $P_{a|i}=(\openone+a \sigma_i)/2$ introduce the two-qubit effects
\begin{equation}
   \tilde{P}_{a|i}:= \frac{1}{2}\qty[P_{a|i}\otimes \openone + \openone \otimes \Phi(P_{a|i})].
\end{equation}
Using $\Phi(P_{k|i})= \qty(\openone - a \sigma_i)/2$, we obtain
\begin{equation}\label{effopK}
 \tilde{P}_{a|i}=\frac{1}{2} \openone_4 + \frac{a}{4}K_i.   
\end{equation}
These effects form a binary POVM and satisfy $\Tr[\rho^{\rm ap} \tilde{P}_{a|i}]=(1+a m_i)/2=\Tr[\rho(\vec m)P_{a|i}]$. It therefore suffices to construct a parent POVM with operator marginals
\begin{equation}
    \sum_{b,c}\Pi_{abc}=\tilde{P}_{a|x}, \quad \sum_{a,c}\Pi_{abc}=\tilde{P}_{b|y}, \quad \sum_{a,b}\Pi_{abc}=\tilde{P}_{c|z}.
\end{equation}
To this end, define $S_{ij}:=\sigma_i\otimes \sigma_j + \sigma_j\otimes \sigma_i$, $i\neq j$, and consider the ansatz
\begin{equation} \label{jpovmxyzans}
    \Pi_{abc}=\alpha \openone_4 + a q_x K_x + b q_y K_y + c q_z K_z  + ab q_{xy} S_{xy}+ bc q_{yz} S_{yz}+ ca q_{zx} S_{zx}, \quad \alpha,q_i,q_{ij}\in \mathbb{R}.
\end{equation}
The terms proportional to $K_i$ determine the individual marginals, while the terms proportional to $S_{ij}$ vanish under the corresponding marginal sums and can be used to ensure positivity. This is a restricted ansatz for constructing a parent POVM, rather than a parametrisation of every possible joint measurement.

Summing over all outcomes eliminates every term except the identity contribution. Normalisation consequently gives $\sum a = \sum (ab)=0$, and $\sum_{abc}\Pi_{abc}=8\alpha\openone_4$, implying $\alpha=1/8$. 

Similarly, summing over two indices (lets $b,c$) over the $\Pi_{abc}$ yields
\begin{equation}\label{marprodef}
    \sum_{b,c}\Pi_{abc} = \frac{1}{2}\openone_4 + 4 a q_x K_x. 
\end{equation}
So, comparing Eqs.~\eqref{effopK} and \eqref{marprodef}, together with the corresponding conditions for the other two marginals, yields
\begin{equation}
    q_x=q_y = q_z = \frac{1}{16}.
\end{equation}
Now, the values of $q_{ij}$ is fixed by the positivity condition $\Pi_{abc}\geq 0$. For simplicity lets assume $q_{xy}=q_{yz}=q_{zx}=q$. If there exists a valid region of $q_{ij}$ for all $i\neq j$, then suitable choice of $q_{ij}$ in that region will be sufficient for positivity of $\Pi_{abc}$. Now, solving the eigenvalue equation, we find that the four eigenvalues are
\begin{equation}
    \mu_1=\mu_2=\frac{1}{8}+2q, \quad \mu_{\pm}=\frac{1}{8}\qty(1-16q \pm \sqrt{3+256 q^2}).
\end{equation}
The conditions $\mu_1,\mu_2\geq0$ require $q\geq-1/16$. Meanwhile, $\mu_-\geq0$ implies $1-16q \geq \sqrt{3+256q^2}$. Squaring this inequality gives $q\leq-1/16$. Hence, within the symmetric one-parameter ansatz, positivity uniquely determines $q=-1/16$. At this value, $\rm spec \ \Pi_{abc}=\{1/2,0,0,0\}$. The parent POVM is therefore
\begin{equation} \label{jpovmxyz}
    \Pi_{abc}=\frac{1}{16} \qty[2 \openone_4 + a K_x + b K_y
    + c K_z - ab S_{xy} -bc S_{yz} - ca S_{zx}], \quad a,b,c \in \qty{\pm 1}. 
\end{equation}
Every effect is positive semidefinite and rank one, and the eight effects sum to $\openone_4$. Their marginals are precisely the $\tilde P_{a|i}$. This recovers the joint measurement constructed in \cite{Patra2026}. The corresponding outcome probabilities are
\begin{equation}
    p(a,b,c|\vec m)=\frac{1}{8}\qty(1+am_x+bm_y+cm_z+abm_xm_y+bcm_ym_z+cam_zm_x).
\end{equation}
Summation over any two outcomes gives the required single-qubit Pauli statistics. Thus, a single measurement on the antiparallel pair produces three classical outcomes with the prescribed marginals. This statement concerns the family $\rho(\vec m)\otimes\rho(-\vec m)$; it does not imply joint measurability of the sharp Pauli observables on a single qubit.

The relation to the metrological result follows from the operator identity $[K_i,K_j]=2\iota \sum_k \epsilon_{ijk}J_k$. Although the difference operators do not commute, their mean commutators vanish $\Tr[\rho^{\rm ap}[K_i,K_j]]=0$. For a parameter-dependent qubit state $\rho_{\vec\theta}$, with single copy SLD $L_\mu=\alpha_\mu\openone_2+\vec{l}_\mu\cdot\vec{\sigma}$, with $\alpha_\mu,\vec{l}_\mu\in \mathbb{R}$. As established in Sec. III, the spin-flipped family admits the SLD $L_\mu^{\Phi}=\alpha_\mu\openone_2-\vec{l}_\mu\cdot\vec{\sigma}$, the collective antiparallel SLD is thus 
\begin{equation}
    L_\mu^{\rm ap}=2\alpha_\mu\openone_4+\vec{l}_\mu\cdot\vec{K}.
\end{equation}
Its commutator satisfies $[L_\mu^{\rm ap},L_\nu^{\rm ap}]=2\iota\qty(\vec{l}_\mu\times \vec{l}_\nu)\cdot\vec{J}$. Since the antiparallel family has $\expval{\vec J}=0$, $\Tr[\rho_\theta^{\rm ap}[L_\mu^{\rm ap},L_\nu^{\rm ap}]]=0$. This identifies the algebraic cancellation underlying weak compatibility without requiring commutativity of the SLDs.

The joint-measurement construction and the metrological cancellation nevertheless impose different requirements. The former depends on positivity of a parent POVM with prescribed marginals; the latter concerns the mean commutators of the SLDs. In particular, the parent POVM above need not extract all the available SLD-FIM.

To illustrate this distinction explicitly, consider the three-parameter qubit model $\rho(\vec{m})=(\openone+\vec{m}\cdot\vec\sigma)/2$, with the Bloch-vector components $(m_x,m_y,m_z)$ as the unknown parameters. The antiparallel configuration is $\rho^{\rm ap}=\rho(\vec{m})\otimes (-\vec{m})$. We evaluate its metrological properties at $\vec m=\emptyset$. At this point $\rho^{\mathrm{ap}}(\emptyset)=\openone_4/4$, and its SLD operators are $L_i^{\mathrm{ap}}=K_i$. Therefore, the SLD-FIM is
\begin{equation}
 \qty[F_{\mathrm{SLD}}^{\mathrm{ap}}(\emptyset)]_{ij}=\Tr[\rho^{\rm ap}(\emptyset)K_iK_j] =2 \delta_{ij} \implies   F_{\mathrm{SLD}}^{\mathrm{ap}}(\emptyset)=2I_3.
\end{equation}
Since the mean SLD commutators vanish, this matrix characterises the asymptotically attainable SLD-CRB.

On the other hand, the joint probabilities generated by the parent POVM satisfy $p(a,b,c|\emptyset)=1/8$, with derivatives $\partial_{m_x}p=a/8$, $\partial_{m_y}p=b/8$ and $\partial_{m_z}p=c/8$. The resulting OFIM is $F_{\mathrm C}(\emptyset)=I_3$. Thus, $F_{\mathrm C}(\emptyset)=\frac12F_{\mathrm{SLD}}^{\mathrm{ap}}(\emptyset)$.

This explicitly demonstrates that reproducing the statistics of three incompatible observables through a common parent POVM does not necessarily yield an optimal multiparameter estimation strategy. In contrast, weak compatibility ensures the asymptotic attainability of the SLD bound through suitable collective measurements on independent repetitions of the antiparallel configuration. The two notions of compatibility are therefore operationally distinct, notwithstanding their common algebraic origin in the antiparallel spin structure.


\end{document}